\documentclass{ws-p8-50x6-00}
\def\ifmath#1{\relax\ifmmode #1\else $#1$\fi}

\def\brk{{\rm {\bf k}}}
\def\dst{\displaystyle{\phantom{|}}}
\def\ov{\over}

\def\be{\begin{equation}}
\def\ee{\end{equation}}
\def\bea{\begin{eqnarray}}
\def\eea{\end{eqnarray}}

\begin{document}

\title{Coulomb and core/halo corrections to Bose-Einstein 
	$n$-particle correlations}

\author{E. O. ALT}

\address{Institut f\"ur Physik, Universit\"at Mainz, 
	D - 55099 Mainz, Germany \\
    E-mail: Erwin.Alt@uni-mainz.de}

\author{T. CS\"{O}RG\H{O}}

\address{MTA KFKI RMKI, H-1525 Budapest 114, POB 49, Hungary \\
    E-mail: csorgo@sunserv.kfki.hu}

\author{B. L\"ORSTAD}

\address{Kristianstad University, S - 29188 Kristianstad, Sweden \\
    E-mail: bengt.lorstad@cf.hkr.se }

\author{J. SCHMIDT-S{\O}RENSEN}

\address{Physics Department, University of Lund, S - 221 00 Lund, Sweden \\
    E-mail: sorensen@alf.nbi.dk}


\maketitle

\abstracts{
We report on a systematic treatment of
Coulomb corrections for 3- and $n$-particle (Bose-Einstein)
correlations, leading to a removal of a 100 \% systematic
error from earlier Coulomb corrections of 5-particle
Bose-Einstein correlations in heavy ion collisions.
These results are generalized to include strong 
final state interactions and core-halo effects.
}

\section{Introduction}

One of the most important tasks of high energy heavy-ion 
studies is to prove the existence of the elusive quark-gluon 
plasma and to study the properties of this predicted new 
state of matter\cite{stock}. Hanbury-Brown Twiss (HBT) 
interferometry\cite{hanbury} of identical particles has 
become an important tool as it can be used to measure the 
evolving geometry of the interaction region,
see refs.\cite{Boa90a,Lorstad,uw-rev,w-rev,cs-revs} 
for some of the recent reviews of this rapidly expanding field.
The quantitative interpretation of the HBT-results depends 
critically on the understanding
of the r{\^o}le of the Coulomb 
interaction between the selected particles. 
However, Coulomb interactions between three (or
more) charged particles are notoriously difficult to handle, 
as evidenced by the many decades of research 
in atomic physics aimed at obtaining solutions of the 
three-body Coulomb scattering problem which are 
accurate and possess a wide range of applicability.

Starting from a three-body Coulomb wave function which
is at least exact for sufficiently large inter-particle separations, 
we determined the effect of Coulomb final state interactions 
on the three-particle Bose-Einstein correlation function 
of similarly charged particles in ref.\cite{c3}. 
We estimated numerically that the 
familiar Riverside approximation is not precise enough to 
determine the three-body Coulomb correction factor
in the correlation function if the characteristic HBT 
radius parameter is between 5 and 10 fm,
which is the range of interest in high-energy heavy ion physics.

We then generalized the method of Coulomb wave function corrections
from three to $n$ charged particles in ref.\cite{cn}.
Here, we re-formulate this work so as to be able to
include the possible effects of 
i) additional strong final state interactions 
and ii) core-halo model corrections.

\section{Coulomb and final state interaction effects in
$n$-particle correlations} 

The $n$-particle Bose-Einstein correlation function is defined as
\begin{equation} 
	C_n(\brk_1, \cdots , \brk_n ) = {\dst N_n(\brk_1, \cdots , \brk_n) \ov
		N_1(\brk_1) \cdots N_1(\brk_n) },
\end{equation}
where $N_n(\brk_1, \cdots , \brk_n) $ is the $n$-particle, and $N_1(\brk_i)$ 
the single-particle inclusive invariant momentum distribution. 
The three-momentum vector of particle $i$ is denoted by 
$\brk_i$. It is quite remarkable that this complicated 
object, that carries quantum
mechanical information on the phase-space distribution of particle 
production as well as on possible partial coherence of the source,
can be expressed in a relatively simple, straight-forward manner
both in the analytically solvable pion-laser model of 
refs.\cite{pratt,cstjz,jzcst} as well as 
in the generic boosted-current formalism of 
Gyulassy and Padula\cite{gyu-pa} as
\begin{equation}
	C_n(\brk_1, \cdots , \brk_n ) = {\dst
		\sum_{\sigma^{(n)}} \prod_{i = 1}^n G(\brk_i,\brk_{\sigma_i})	
		\ov \prod_{i=1}^n G(\brk_i,\brk_i) }, \label{e:m2}
\end{equation}
where $\sigma^{(n)}$ stands for the set of permutations of indices
$(1, 2, \cdots, n)$ and $\sigma_i $ denotes that element which replaces
element $i$ in a given permutation from the set of $\sigma^{(n)}$. 
Regardless of the details of the two different derivations
\begin{equation}
 G(\brk_i,\brk_j) = \langle a^{\dagger}(\brk_i) a(\brk_j) \rangle
	\label{e:bas0}
\end{equation}
stands for the expectation value of $a^{\dagger}(\brk_i) a(\brk_j) $.

In the relativistic Wigner-function formalism, 
in the plane wave approximation $G(\brk_1,\brk_2)$ 
can be rewritten as
\begin{eqnarray}
	G(\brk_1,\brk_2) & = & \int d^4 x \,S(x,K_{12}) \,\exp(i q_{12}\cdot x)
		\label{e:gwig}\\
	K_{12} & = & 0.5 (k_1 + k_2), \qquad q_{12} \, = \, k_1 - k_2,
\end{eqnarray}
where a four-vector notation is introduced,
$k_i = (\sqrt{m_i^2 + \brk_i^2}, \brk_i)$, and $a \cdot b$ 
stands for the 
inner product of four-vectors.
Due to the mass-shell 
constraints $E_{\brk_i} = \sqrt{m^2_i + \brk^2_i}$,
$G$ depends only on 6 independent momentum components.
In any given frame, the boost-invariant decomposition
of Eq.~(\ref{e:gwig}) can be rewritten in the following,
seemingly not invariant form:
\begin{eqnarray}
	G(\brk_1,\brk_2) & = & \int d^3 {\bf x}\,\, S_{{\bf K}_{12}}({\bf x})\,\,
		\exp( - i {\bf q}_{12} {\bf x}),
		\label{e:gwignr}\\
	S_{{\bf K}_{12}}({\bf x}) 
	& = & \int dt\, \exp( i {\mbox {\boldmath $\beta$}}_{K_{12}} 
	{\bf q}_{12} t)\,\, S( {\bf x}, t, K_{12}), \\
	{\mbox {\boldmath $\beta$}}_{K_{12}} & = & (\brk_1 + \brk_2) /(E_1 + E_2).
\end{eqnarray}
	If $n$ particles are emitted with similar momenta
	so that their $n$-particle Bose-Einstein correlation
	functions may be non-trivial, Eqs.~(\ref{e:m2},\ref{e:bas0})
	will form the basis for evaluation of 
	the Coulomb and strong final state interaction effects 
	on the observables.
	On this level, all the correlations are build up from correlations
	of pairs of particles.

	In order to treat the Coulomb corrections to the 
$n$-particle correlation function exactly, knowledge 
of the $n$-body Coulomb scattering wave 
function is required. We restrict ourselves to the case 
that the transverse momenta of all
$n$ particles in the final state in their common center of mass are 
small enough to make a nonrelativistic approach sensible. 
Hence the problem consists in finding 
the solution of the $n$-charged particle Schr\"odinger 
equation when all $n$ particles are in the continuum. 

Consider $n$ distinguishable 
particles with masses $m_{i}$ and charges 
$e_{i},\,i = 1,2,$ $\cdots $, $n$. Let ${\rm {\bf {x}}}_{i}$ 
denote the coordinate (three-)vector of particle $i$. Then 
${\rm {\bf {r}}}_{ij} = {\rm {\bf {x}}}_{i} - 
{\rm {\bf {x}}}_{j}$ is the relative coordinate 
between particles $i$ and $j$, and ${\rm {\bf {k}}}_{ij} = 
(m_j {\rm {\bf {k}}}_{i} - m_i {\rm {\bf {k}}}_{j})/
{(m_{i} +m_{j})}$ the canonically 
conjugate relative momentum. 

The $n$-particle Schr\"odinger equation reads as
\begin{eqnarray}
\left\{ H_0 + \sum_{i < j=1}^{n} V_{ij} - E \right\} 
\Psi^{(+)}_{ {\rm {\bf {k}}}_{1} \cdots {\rm {\bf {k}}}_{n} }
({\rm {\bf {x}}}_{1},\cdots, {\rm {\bf {x}}}_{n}) = 0, 
\label{nse}
\end{eqnarray}
where
\begin{eqnarray}
E = \sum_{i=1}^{n}\frac{{\rm {\bf {k}}}_{i}^2}{2m_{i}} > 0
\end{eqnarray}
is the total kinetic energy for $n$ particles in the continuum. 
$H_0$ is the free Hamilton operator and
\begin{eqnarray}
V_{ij}({\rm {\bf {r}}}_{ij}) = V_{ij}^S({\rm {\bf {r}}}_{ij}) 
+ V_{ij}^C({\rm {\bf {r}}}_{ij}) 
\end{eqnarray}
the interaction between particles $i$ and $j$, 
consisting of a short-range strong potential $V_{ij}^S$ and the 
long-range Coulomb interaction, $V_{ij}^C({\rm {\bf {r}}}_{ij}) 
= e_i e_j/\mid {\rm {\bf {r}}}_{ij} \mid$. 

Although for $n= 3$ an exact numerical solution of the 
Schr\"odinger equation (\ref{nse}) with $V_{ij} = V_{ij}^C$
has been achieved recently for $E>0$\cite{brilm01}, 
because of its complexitiy the resulting wave function can not 
easily be used for general purposes. In addition, 
for $n>3$ a numerical solution 
is beyond present means. For a brief discussion of 
the related difficulties see\cite{a98b}. This means 
that for practical purposes analytic, even if only 
approximate, $n$-charged particle wave functions are 
still needed. 

For $n=3$ such wave functions are available in the form 
of the explicit
solutions of the Schr\"odinger equation in all asymptotic 
regions of the three-particle configuration space\cite{r72,am92}. 
The simplest form, which is also easily generalized to arbitrary 
particle numbers, applies to the (dominant) asymptotic region 
conventionally denoted by $\Omega _{0}$ and characterized 
by the fact that - roughly speaking - all three inter-particle 
distances become uniformly large\cite{am92}. 
In the final states of heavy-ion
reactions, where a large number of charged particle tracks appear, 
the mutual, macroscopically large separation of tracks
is one of the criteria of a clean measurement. This suggests 
that in order to study Coulomb effects on $n$-body correlation 
functions, knowledge of the wave function in 
$\Omega_{0}^{(n)}$, the region in $n$-particle configuration 
space where all interparticle distances become uniformly 
large, i.e., $\mid {\rm {\bf {r}}}_{ij} \mid 
\to \infty$ for all values of $(ij)$, may be sufficient. 
\begin{figure}[t]
\centerline{\epsfxsize=15pc 
\epsfbox{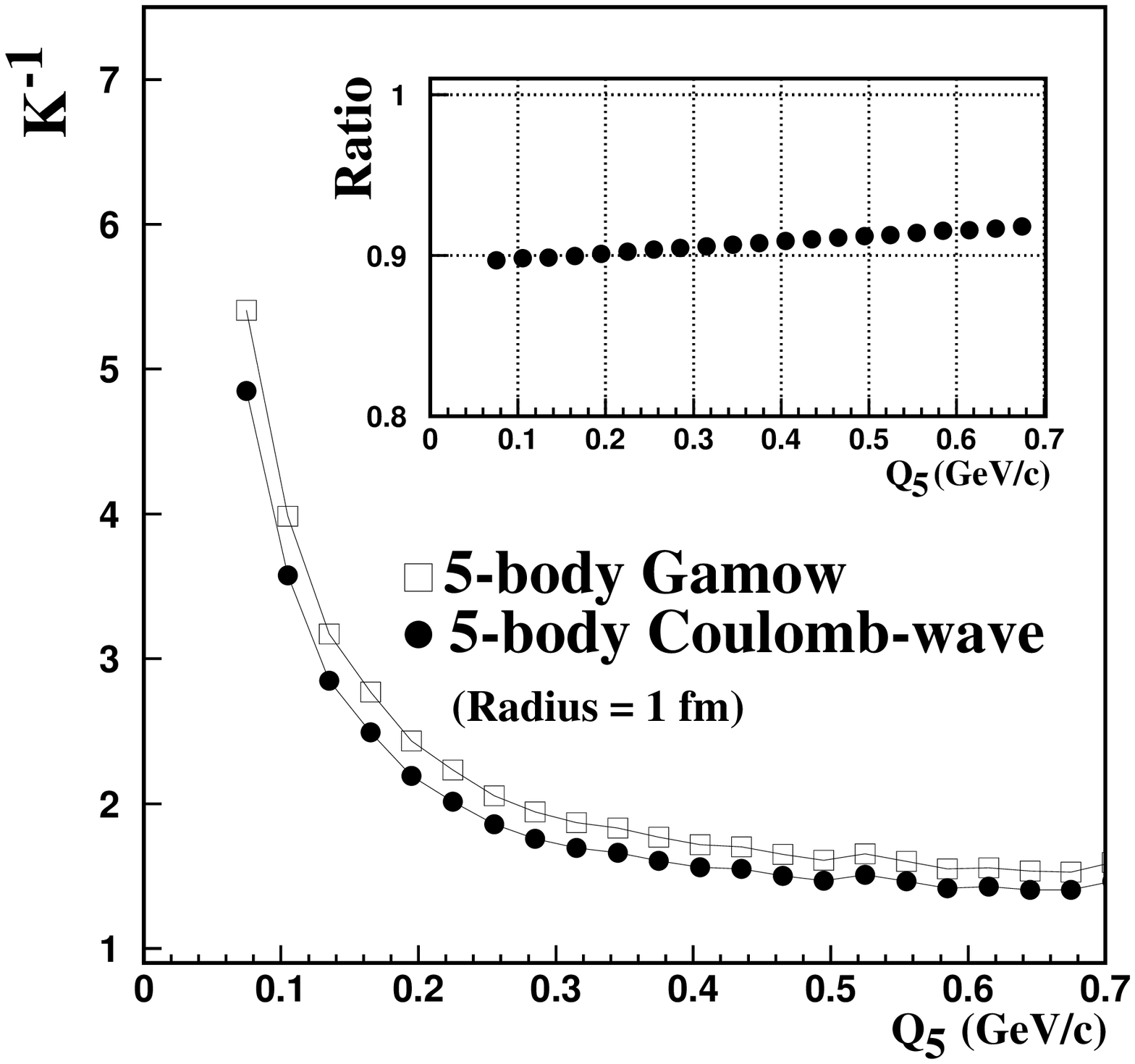}
\epsfxsize=15pc 
\epsfbox{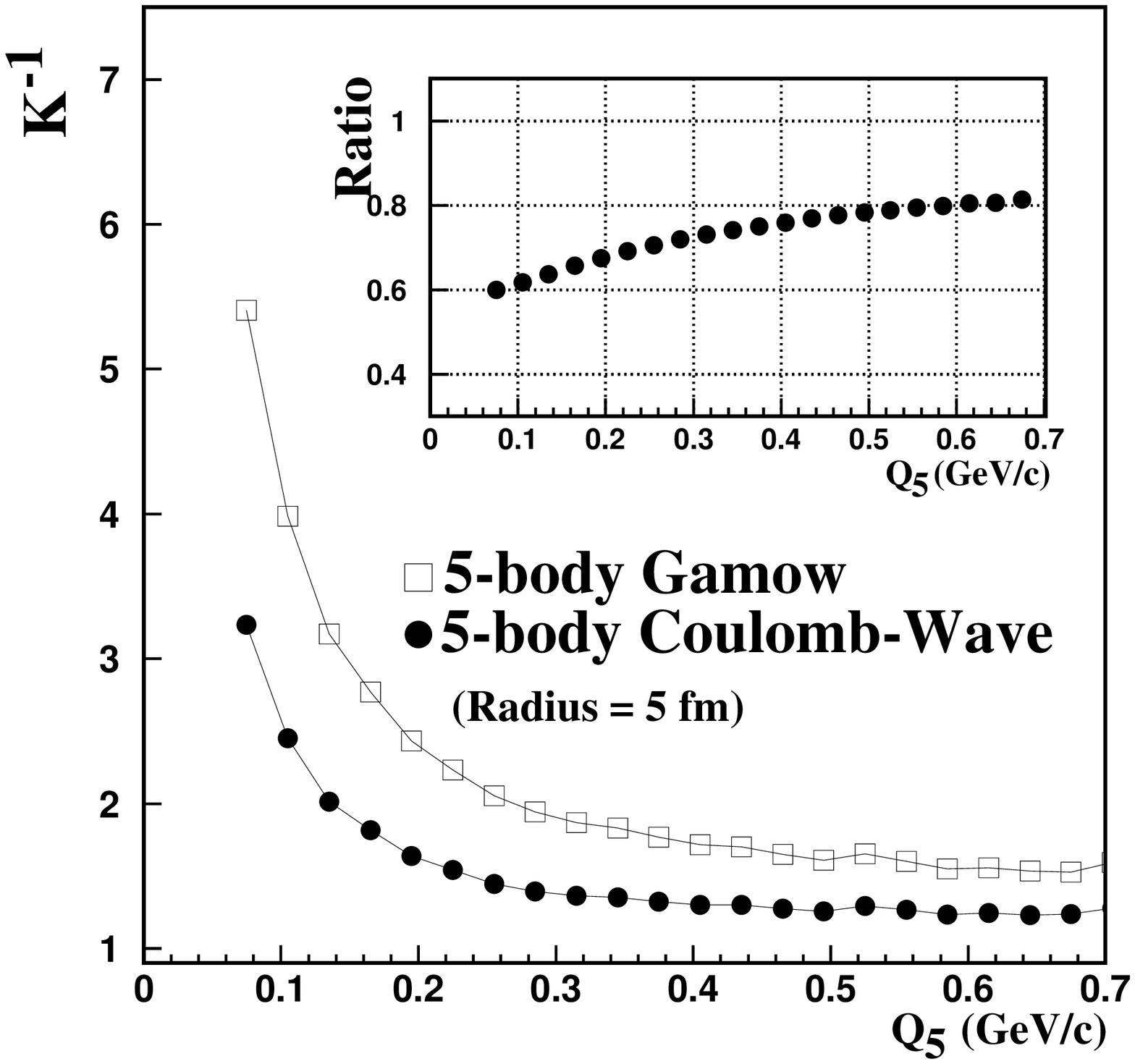}} 
\vspace{-0.5 truecm}
\caption{Coulomb wave function correction (filled circles) as 
compared to the less substantiated Gamow correction (squares) 
for 5-particle Bose-Einstein correlations for a source sizes 
$R= 1 $ fm (left panel) and $R=5 $ fm (right panel).
	The inset shows the ratio of these two correction factors.
\label{f:1}}
\end{figure}
Hence, an appropriate, approximate $n$-particle Coulomb 
scattering wave function is
\begin{eqnarray} 
\Psi^{(+)}_{ {\rm {\bf {k}}}_{1}, \cdots ,
 {\rm {\bf {k}}}_{n} }({\rm {\bf {x}}}_{1}, \cdots ,{\rm {\bf {x}}}_{n}) \; 
\sim \; \prod_{i < j=1}^n 
{\psi}_{ {\rm {\bf {k}}}_{ij}}^{C(+)}({\rm {\bf {r}}}_{ij}).
\label{aswn}
\end{eqnarray} 
Here, ${\psi}_{ {\rm {\bf {k}}}_{ij}}^{C(+)}({\rm {\bf {r}}}_{ij})$ 
is the continuum solution of the two-body 
Coulomb Schr\"odinger equation 
(with a reduced mass of $\mu_{ij} = m_{i}m_{j}/(m_{i} +m_{j})$):
\medskip
\begin{eqnarray} 
\left\{ - \frac{\Delta_{{\rm {\bf{r}}}_{ij}}}
{2 \mu_{ij}} + V_{ij}^C({\rm {\bf{r}}}_{ij}) - \frac {{\rm {\bf {k}}}_{ij}^{2}}
{2 \mu_{ij}} \right\} 
{\psi}_{ {\rm{\bf{k}}}_{ij}}^{C(+)}({\rm {\bf{r}}}_{ij}) = 0, 
\label{2cse} 
\end{eqnarray}
describing the relative motion of the two particles $i$ and $j$ with 
energy ${\rm {\bf {k}}}_{ij}^{2}/2 \mu_{ij}$. The explicit solution is 
\medskip
\begin{eqnarray} 
	{\psi}_{ {\rm {\bf {k}}}_{ij}}^{C(+)}
	({\rm {\bf {r}}}_{ij}) 
	&=& N _{ij} \; e^{ i {\rm {\bf {k}}}_{ij} 
	{\rm {\bf {r}}}_{ij}} 
	F[- i \eta _{ij}, 1; i ( \mid {\rm {\bf {k}}}_{ij} \mid 
\mid {\rm {\bf {r}}}_{ij} \mid - {\rm {\bf {k}}}_{ij} 
	{\rm {\bf {r}}}_{ij})], \label{2pwfn}
\end{eqnarray} 
with $N_{ij} = e^{- \pi \eta_{ij}/2}\, 
\Gamma (1 + i \eta_{ij}), $ and $\eta_{ij}= 
e_{i} e_{j}\mu_{ij}/ \mid {\rm {\bf {r}}}_{ij} \mid $ being the 
Coulomb parameter. $F[a,b;x]$ is the confluent hypergeometric 
function and $\Gamma (x)$ the Gamma function. The foundation of 
Eq.~(\ref{aswn}) in fundamental scattering theory and its 
expected range of validity are discussed in ref.\cite{cn}.

	Additional strong final state interactions can be taken into account in
	a straight-forward manner by substituting in Eq.~(\ref{aswn}) 
	for the two-particle Coulomb wave functions 
${\psi}_{ {\rm{\bf{k}}}_{ij}}^{C(+)}({\rm {\bf{r}}}_{ij})$ 
the solutions of the Schr\"odinger equation with the Coulomb 
plus strong potential,
\begin{eqnarray} 
	\left\{ - \frac{\Delta_{{\rm {\bf{r}}}_{ij}}}
	{2 \mu_{ij}} + V_{ij}^C({\rm {\bf{r}}}_{ij}) 
	+ V_{ij}^S({\rm {\bf{r}}}_{ij}) 
	- \frac {{\rm {\bf {k}}}_{ij}^{2}}
	{2 \mu_{ij}} \right\} 
	{\psi}_{ {\rm{\bf{k}}}_{ij}}^{CS(+)}({\rm {\bf{r}}}_{ij}) = 0. 
	\label{2csse} 
\end{eqnarray}
	In this way a few dominant, strong interaction 
induced phase-shifts can be taken into account
for the pair under consideration. On the
level of the two-particle correlation function,
a similar technique has already been applied in ref.\cite{pratt-cs}.

\section{Application to high-energy heavy-ion and particle collisions}

The correlation function measuring the enhanced probability for emission of 
$n$ identical Bose particles is given by Eq.~(1).
 This correlation function
is usually, due to meager statistics, 
only measured as a function of the Lorentz 
invariant $Q_n$, defined by the relation
\begin{equation}
Q_n^2= \sum_{i<j = 1}^n ({k}_i-{k}_j)^2.
\end{equation}

We can now calculate the Coulomb effects on 
the $n$-particle correlation function using 
\begin{equation}
K_{Coulomb}(Q_{n})= 
\frac{ \int \prod_{i = 1}^n d^{3}{\bf x_i} \rho ({\bf x_i})
\left|\Psi^{(+){\cal S}}_{ {\rm {\bf {k}}}_{1} 
	\cdots {\rm {\bf {k}}}_{n} }
	({\rm {\bf {x}}}_{1},\cdots,
		{\rm {\bf {x}}}_{n})\right|^2}
	{ \int \prod_{i = 1}^n d^{3}{\bf x_i} \rho ({\bf x_i})
\left|\Psi^{(0){\cal S}}_{ {\rm {\bf {k}}}_{1} 
	\cdots {\rm {\bf {k}}}_{n} }
({\rm {\bf {x}}}_{1},\cdots, {\rm {\bf {x}}}_{n})\right|^2}.
\label{Kcoul3}
\end{equation}
Here, $\rho({\bf x_i})$ is the density distribution 
of the source for particle $i$ (normalized to the total number of particles), 
taken as a Gaussian distribution of width R in 
all three spatial directions and 
$\Psi^{(0)}_{ {\rm {\bf {k}}}_{1} \cdots 
{\rm {\bf {k}}}_{n} }({\rm {\bf {x}}}_{1},\cdots, 
{\rm {\bf {x}}}_{n}) \sim \prod_{i < j=1}^n 
e^{ i {\rm {\bf {k}}}_{ij} {\rm {\bf {r}}}_{ij}}$ 
is the $n$-body wave function 
without any final state interaction. The superscript ${\cal S}$ 
indicates appropriate symmetrisation for identical particles. 
This formulation makes it possible to 
extract information on the source size $R$, 
and to compare this value with that extracted by means of
a generalized $n$-particle Gamow approximation 
via $K_{Coulomb}^{(G)}(Q_{n})= \prod_{i<j = 1}^n G_{ij}$, 
where $ G_{ij}= \mid N _{ij}\mid^2$. 

Pion $n$-tuples were sampled randomly\cite{c3,cn} 
from the NA44 data sample of three pion events produced 
in S-Pb collisions at CERN\cite{janus-na44}. 
Fig. 1 demonstrates that in case of a characteristic effective 
source size of 1 fm, the difference between the 
$n$-particle Gamow and the Coulomb wave function 
corrections are smaller than 10 \% for $n = 5$ particles.
However, with increasing number of particles and/or 
with increasing effective source sizes, this difference 
increases dramatically. For instance, for 5 particles 
the naive generalized Gamow method overestimates the 
Coulomb correction as compared to the much better 
substantiated Coulomb wave function integration method 
by a factor of 2.

\section{Core-halo corrections}

	The results given in the previous section 
	have to be corrected for the effects that may arise from
	the existence of a halo of long-lived hadronic 
	resonances\cite{chalo,biya-c3}. In this case, the effective source
	function $\rho(x)$ has two components, one pertinent to 
	a smaller region
	corresponding to the core of the interactions and
 one to
	a larger region that corresponds to the (unresolvable)
	halo:
\begin{eqnarray}
	\rho(x) & =& \rho_c(x) + \rho_h(x) \\
	\langle n \rangle & = & \langle n_c \rangle + \langle n_h \rangle \\
	\langle n_i \rangle & = & \int dx \rho_i(x), \quad i = c, h.
\end{eqnarray}
	
	The effective intercept parameter of the two-particle 
Bose-Einstein correlation function reads as
\begin{equation}
	\lambda_* = (\langle n_c\rangle /\langle n \rangle)^2
\end{equation}
if the core fraction is independent of the relative momentum.
 If the core-halo model is applicable to a given data set, 
the radius $R_h$ of the halo does not matter as long as 
it is big enough. For instance, it can be a Gaussian with 
a radius of 20-40 fm; in any case $R_h >> \hbar/Q_{min}$, the inverse 
of the two-particle relative momentum resolution.

\section{Summary}
In this contribution we reviewed the state-of-art method
for treating Coulomb corrections to 3- and $n$-body 
correlation functions. 	For small effective source 
sizes of about 1 fm, the pairwise product of Gamow factors 
gives a good approximation for the $n$-body Coulomb 
correction factor. However, for source sizes typical in 
high energy heavy ion experiments, 	a more substantiated 
calculation based on the pairwise product of relative 
Coulomb wave functions is required in order to remove a 
possible 100 \% systematic error from this correction.

	We have also generalized this correction method for the 
	inclusion of strong final state interactions and pointed out that
	the Coulomb corrections have to be performed self-consistently
	with the core-halo corrections;  in particular the
	effective intercept parameter of the two-particle Bose-Einstein
	correlation function has to be taken into account when
	evaluating the 3- or $n$-body Coulomb corrections.

\section*{Acknowledgments}
This research has been supported by
a Bolyai Fellowship of the Hungarian Academy of Sciences,
by the Hungarian OTKA grants T024094, T026435, T029158, T034269,
by the US-Hungarian Joint Fund MAKA 652/1998, by the NWO-OTKA grant
N025186 and by  the Swedish Research Council.

\end{document}